\documentclass[
reprint,
superscriptaddress,
showpacs,
 amsmath,
 amssymb,
 aps,
pre,
floatfix,
]{revtex4-2}
\usepackage{mwe}
\usepackage{graphicx}
\usepackage{dcolumn}
\usepackage{bm}
\usepackage{hyperref}
\usepackage{color}
\usepackage[normalem]{ulem}
\usepackage{hyperref}
\usepackage{orcidlink}

\usepackage{xcolor}
\definecolor{salmon}{rgb}{0.98, 0.5, 0.45}

\begin{document}

\title{Hierarchical organization of critical brain dynamics}
\author{Gustavo G. Cambrainha \orcidlink{0009-0000-8153-305X} }
\affiliation{Departamento de F{\'\i}sica, Centro de Ciências Exatas e da Natureza, Universidade Federal de Pernambuco,
Recife, PE, 50670-901, Brazil}
\author{Daniel M. Castro \orcidlink{0000-0002-2761-2133} }
\affiliation{Departamento de F{\'\i}sica, Centro de Ciências Exatas e da Natureza, Universidade Federal de Pernambuco,
Recife, PE, 50670-901, Brazil}

\author{Leonardo~L.~Gollo~\orcidlink{0000-0003-3505-9259}}
\affiliation{Institute for Cross-Disciplinary Physics and Complex Systems, IFISC (UIB–CSIC), University of the Balearic Islands, Palma de Mallorca, Spain}
\affiliation{Brain Networks and Modelling Laboratory and The Turner Institute for Brain and Mental Health, Monash University, Melbourne, VIC, Australia}

\author{Pedro V. Carelli \orcidlink{0000-0002-5666-9606}}
\affiliation{Departamento de F{\'\i}sica, Centro de Ciências Exatas e da Natureza, Universidade Federal de Pernambuco,
Recife, PE, 50670-901, Brazil}

\author{Mauro Copelli \orcidlink{0000-0001-7441-2858}}
\email{mauro.copelli@ufpe.br}
\affiliation{Departamento de F{\'\i}sica, Centro de Ciências Exatas e da Natureza, Universidade Federal de Pernambuco,
Recife, PE, 50670-901, Brazil}

\begin{abstract}
The hierarchical organization of the brain is a fundamental structural principle, while brain criticality is a leading hypothesis for its collective dynamics. 
However, the connection between structure and signatures of criticality remains an open question. 
Here, we address this issue by applying  phenomenological renormalization group approaches to large-scale neuronal spiking activity from the mouse visual cortex and hippocampus. 
We find that signatures of criticality are not uniform, but instead vary systematically along the known anatomical hierarchy in both brain systems.
Strikingly, the direction along this gradient is inconsistent across different criticality exponents, revealing a nontrivial, measure-dependent organization: exponents based on static properties point to a gradient in one direction, while the exponent based on dynamic properties points in the opposite direction.
Moreover, the signatures across the visual system are strongly modulated by the engagement in a visual task. 
We show that the correlations among criticality markers of different brain regions during active engagement are sufficient to reconstruct the anatomical hierarchy from the dynamics. 
Scaling exponents closely follow a theoretically predicted scaling relation among them, and covary with the hierarchical position.
Our findings provide a direct link between the collective dynamics of neurons and the macroscopic architecture of the brain.
\end{abstract}
\maketitle

\section{Introduction}

A fundamental challenge in neuroscience is understanding how complex information processing emerges from the many-body dynamics of large-scale neuronal networks.
Given the massive number of interacting components, how do we even begin to look for organizing principles? 
One possibility lies in identifying key macroscopic quantities and their relations.
Powerful hints in this direction emerged from results in two very different experimental setups.
At the turn of the century, Linkenkaer-Hansen \textit{et al.} observed long-range power-law decaying temporal correlations in electroencephalography and magnetoencephalography recordings in humans~\cite{linkenkaer-hansen_long-range_2001}. 
Two years later, the pioneering experimental work of Beggs and Plenz discovered ``neuronal avalanches", which are cascades of spontaneous activity in neural tissue that often follow power-law distributions in their size and duration~\cite{beggs_neuronal_2003}. 
Both results, with drastic differences in spatial and temporal scales between the two experiments, were considered fingerprints of a system poised at a critical point~\cite{chialvo_emergent_2010, munoz_colloquium_2018, TomenHerrmannErnst2019,obyrne_how_2022}. 
These observations gave strength to the brain criticality hypothesis: the idea that the brain maintains itself in the vicinity of a phase transition, theoretically optimizing computation and information processing~\cite{bertschinger_real-time_2004, haldeman_critical_2005, kinouchi_optimal_2006, beggs_criticality_2008, shew_neuronal_2009, shew_information_2011, boedecker_information_2012, gautam_maximizing_2015, avramiea_long-range_2022, barzon_nicoletti_excitation}.

While an increasing body of evidence currently supports this hypothesis~\cite{ribeiro_spike_2010,lombardi_balance_2012,tagliazucchi_criticality_2012,palva_neuronal_2013,shriki_neuronal_2013,cocchi2017criticality,fontenele_criticality_2019,jones_shew_scalefree}, relying solely on specific signatures like power-law distributed avalanches can be ambiguous~\cite{Touboul2017, Carvalho2021subsampled}. 
Over time, new approaches have been developed to explore neural collective behavior from a different perspective.
One such approach came from the application of the renormalization group (RG) framework to neural population data, as proposed by  Bialek and collaborators~\cite{bradde_pca_2017, Meshulam2019coarse}. 
Due to its model-free nature, this method became known as the phenomenological renormalization group (PRG).
In one of its forms, PRG involves systematically coarse-graining the data by excluding its less dominant collective modes of fluctuation. 
The distribution of these coarse-grained variables will, by the central limit theorem, become Gaussian for a trivial system. 
However, for a system at a critical point, strong correlations defy this convergence, resulting in non-Gaussian distributions with heavy tails. 
The kurtoses of these distributions have been used as a simple measure of ``distance to triviality"~\cite{castro_and_2024, cambrainha2025criticality, nasciment2025arXiv}.

Recently, activity in the mouse primary visual cortex was shown to exhibit robust signatures of criticality, with signature strength increasing with task engagement~\cite{cambrainha2025criticality}. 
This naturally raised the question:
is criticality in the primary visual cortex a local feature emerging due to the visual nature of the task, or does it extend to other brain regions? 
If so, to which extent? 
While signatures of scale-invariance have been found across many brain areas, a clear picture of how these dynamics are systematically organized remains highly controversial. 
The literature presents conflicting gradients: while some evidence suggests that distance to criticality increases along the hierarchical levels~\cite{Morales2023scaling}, other findings propose the exact opposite, with the distance decreasing at higher hierarchical levels~\cite{harris2024tracking}.

Anatomically, cortical hierarchies~\cite{wang_2012_newtorkanalysis, bienkowski2018integration, dsouza2022hierarqui} stand out as an immediate candidate for this dynamical organization.
Cortical hierarchies refer to the structure of information flow across different regions of the cerebral cortex, where some areas process more basic or lower-level information (e.g., sensory input), and others perform increasingly complex, abstract, or integrative operations. 
The hierarchical relationships can often be inferred from anatomical features like the pattern of cortical connections, where the specific pattern of feedforward and feedback projections helps define a directional flow of information. Overall, hierarchies are understood as a major organizational principle of the cortex and its computational function~\cite{vezoli2021cortical}.

Despite the importance of this structural principle, its relationship with critical dynamics has only recently begun to be explored~\cite{Morales2023scaling,muller2023spatial,harris2024tracking, Ponce2025network}. 
In most of these studies, temporal correlations were employed to infer closeness to criticality~\cite{muller2023spatial,harris2024tracking}, since longer autocorrelation times are interpreted as stronger signatures of scaling behavior, due to the expected ``critical slowing down'' close to the phase transition. 
Moving from initial, lower-level processing (say, the primary visual cortex) to subsequent, higher-level structures (say, the anteromedial visual cortex), autocorrelation times were observed to increase, a signature of the ``critical slowing down''.
Signatures of criticality relying on the functional connectivity of the data, on the other hand, revealed the reverse trend, increasing as one ascends in the hierarchy~\cite{Morales2023scaling}.

Here we implement a more robust marker for the non-Gaussianity of the coarse-grained activity, based on the Jensen-Shannon divergence relative to a Gaussian distribution, to more accurately estimate distance to triviality. We show that this dynamical signature reflects features of criticality, and can meaningfully represent known hierarchies in the brain: the anatomical hierarchy can be inferred solely from the spatial distribution of these criticality markers. 
The scaling exponents characterizing the dynamics closely obey a theoretically predicted scaling relation proposed from functional magnetic resonance imaging (fMRI) experiments~\cite{castro2025interdependent}, and follow a gradient aligned with the hierarchical organization of anatomical structure. 

\section{Results}\label{results}

\begin{figure*}[ht!]
\centering
\includegraphics[width=0.9\textwidth]{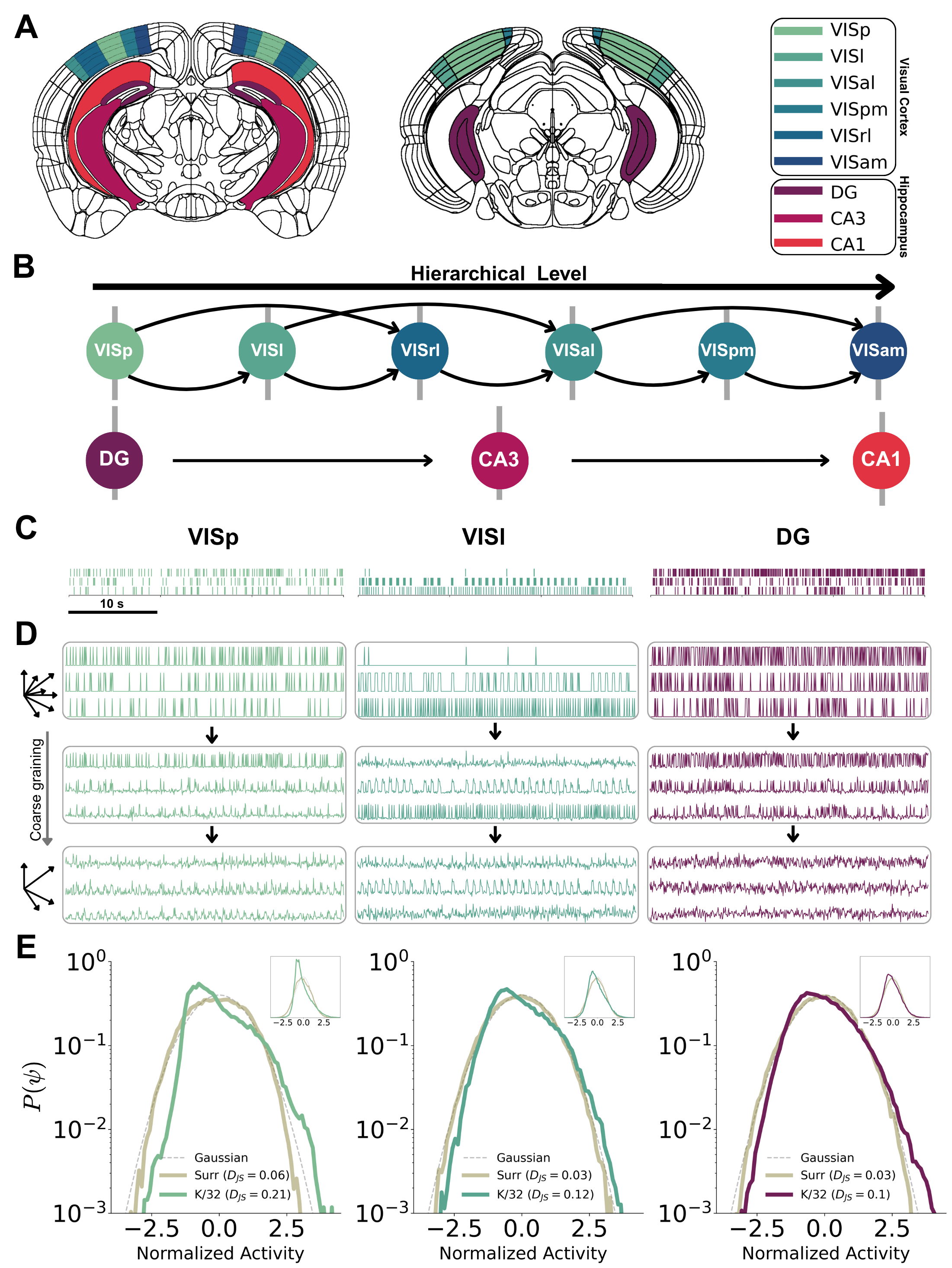}
\caption{(A) Schematic of a mouse brain with the regions considered in this work highlighted. Credit: \href{https://atlas.brain-map.org/atlas}{Allen Institute}. (B) Schematic representation of mouse brain visual hierarchy as present in Harris et al.~\cite{harris2024tracking}
(C) Raster plots of spiking activity during a 30-s window from three brain regions: VISp (left), VISl (center), and DG (right).
(D) Successive coarse-graining of unit time series obtained by projection onto progressively fewer eigenvectors of the covariance matrix ($N$, $N/4$, $N/32$). Variance is normalized to one.
(E) At the final coarse-graining step ($N/32$), brain regions near criticality retain non-Gaussian distributions, whereas surrogate data converge to trivial values. Insets show the same distributions, but with linear vertical axes.}
\label{fig:methods}
\end{figure*}

We analyzed publicly available data from the Allen Visual Behavior Neuropixels dataset~\cite{allendandi}, where the spiking activity of many distinct brain regions was recorded simultaneously while mice performed a visual task (see Methods). 
The experimental session consisted of three primary phases: an ``active" phase, where trained mice were rewarded for licking a spout upon detecting an image change; a ``passive" phase, where the same images were shown but the licking spout was retracted; and an intervening period of non-natural image (NNI) presentation, where images such as Gabor patches, black and white flashes and gray screen are shown to the mice (also without any reward).

Our analyses focused on the population dynamics within individual brain regions. 
Since the phenomenological renormalization group (PRG) method requires a large number of simultaneously recorded units to produce reliable results, we restricted our analysis to brain areas containing more than 128 units in each session~(Fig.~\ref{fig:methods}A). 
In the Methods section, Table~\ref{tabela} displays the acronyms and number of units of each region. 
For each selected session, we applied the PRG method to 30-second windows of neural activity~(Fig.~\ref{fig:methods}C). 
This involved computing the covariance matrix of the $N$ binarized spike trains within each window and extracting its eigenvectors to construct projectors onto their eigenspace~\cite{bradde_pca_2017}. 
The activity of each neuron was coarse-grained by projecting it onto a subspace spanned by a decreasing number of projectors, keeping only the largest $N$, $N/2$, ..., $N/32$ eigenvalues~(Fig.~\ref{fig:methods}D). 
Repeating this procedure for all $N$ neurons, we obtained a probability distribution of  coarse-grained activity $P(\psi)$ for each brain region~(Fig.~\ref{fig:methods}E;  see Methods).

To assess proximity to criticality, we evaluated how the distribution of coarse-grained activity deviates from a Gaussian at the deepest level of coarse-graining (for this dataset, projection onto the subspace spanned by the largest $N/32$ eigenvalues).
Previous studies with anesthetized animals~\cite{castro_and_2024}, awake animals~\cite{munn_multiscale_org, cambrainha2025criticality} and computational models~\cite{nascimento2025phenomenological} have associated proximity to critical dynamics (in their case measured by the kurtosis of $P(\psi)$) and the emergence of non-trivial scaling.

However, this approach is limited by the fact that kurtosis captures only a single moment of the distribution and therefore does not fully characterize deviations from Gaussianity. 
To address this limitation, we adopt a new approach to quantify the deviation from a Gaussian reference, providing a more complete and robust measure of non-Gaussianity.
For each 30-s window, we compute the Jensen-Shannon distance ($D_{JS}$), defined as the square root of the Jensen-Shannon divergence, between $P(\psi)$ and a Gaussian distribution (see Methods).
We also compare the results to those obtained by initially shuffling the spikes of each neuron (surrogate data, see Fig.~\ref{fig:methods}E). 
Since the Gaussian distribution represents a trivial fixed point of the renormalization group flow we can interpret $D_{JS}$ as an effective distance in distribution space from this fixed point, providing a quantitative measure of deviations from trivial scaling behavior.

\begin{figure*}[ht!]
\centering
\includegraphics[width=\textwidth]{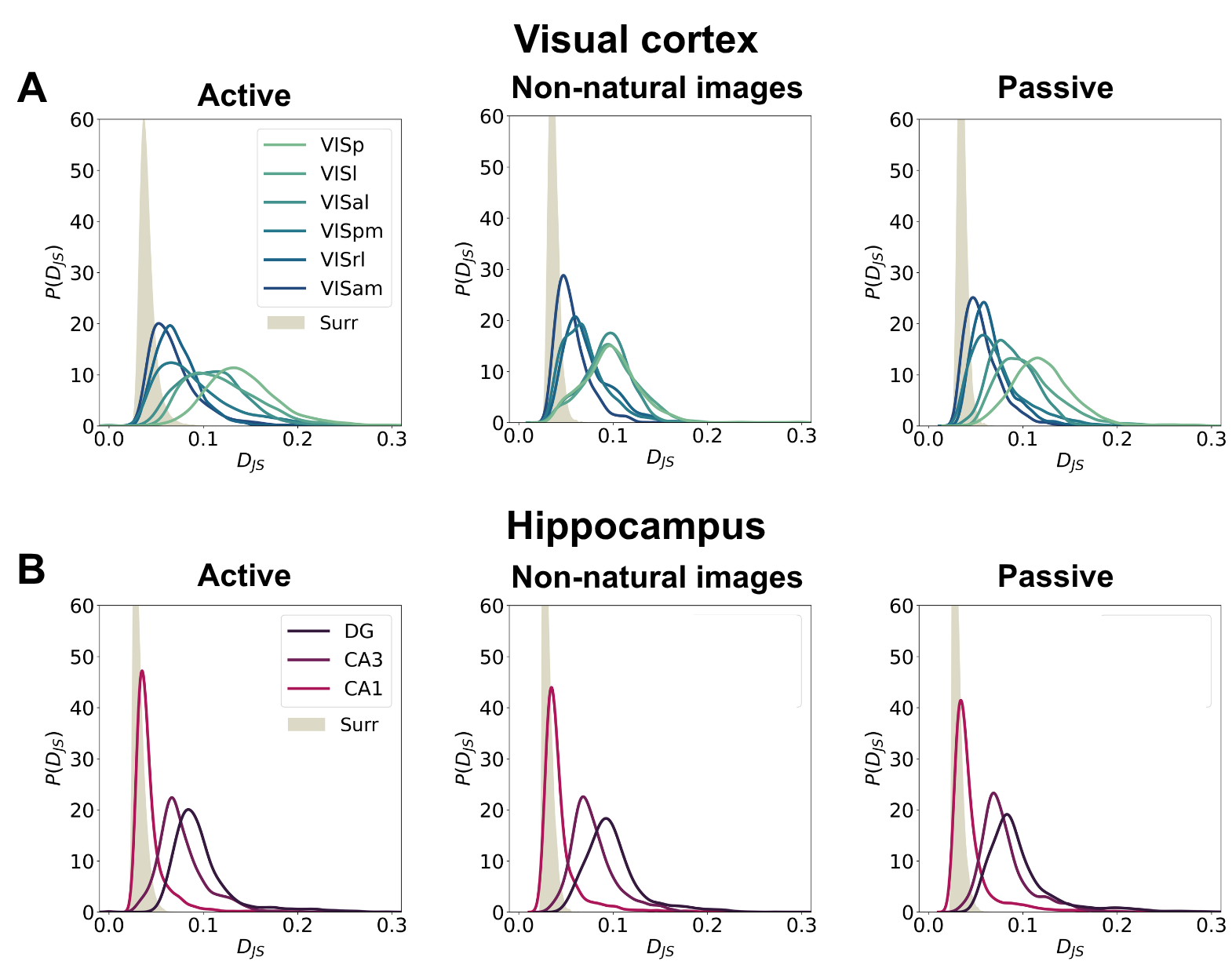}
\caption{Distributions of Jensen-Shannon distance ($D_{JS}$) across different brain regions during different experimental phases.
(A) Distributions of $D_{JS}$ of coarse-grained variables in visual areas to a Gaussian distribution, showing region-specific differences in shape.
(B) Corresponding distributions in hippocampus. Shaded area denotes the distribution of $D_{JS}$ from surrogate data to a Gaussian distribution pooled across all regions in each image.}
\label{fig:kurts}
\end{figure*}

\subsection{The distance to triviality across brain regions}

In the renormalization group framework, only infinite systems poised exactly at criticality can maintain a non-Gaussian form for any amount of coarse-graining~\cite{kadanoff2009more}. 
Real-world systems, however, can only approximate this regime, and are additionally subject to finite-size and noise effects. 
Recent numerical studies of neuronal network models show that coarse-grained activity distributions become increasingly non-Gaussian as the system approaches criticality.
This dependence is smooth, with the deviation from Gaussianity peaking at the critical point~\cite{nasciment2025arXiv}.
To capture this, we used the Jensen-Shannon divergence ($D_{JS}$) between the empirical distributions in a given time window and a Gaussian baseline as a measure of the system’s distance from trivial dynamics (Fig. \ref{fig:methods}D).
Alternative metrics of non-Gaussianity are explored in the Supplementary Material and display similar results.

Since each brain region yields a collection of $D_{JS}$ values, we examined the distribution of these values separately for each region. In the visual cortical areas, $D_{JS}$ distributions exhibit distinct shapes, with noticeable differences in both their peaks and spreads, and most values lie above those obtained from the pooled surrogate distributions (Fig.~\ref{fig:kurts}A). 
These distribution profiles also change across experimental conditions, shifting from active stimulus presentation to non-natural images and, finally, to passive stimulation.
A similar regional variation is observed in the hippocampus, where the strongest deviations from trivial dynamics occur in the Dentate Gyrus. 
However, compared to the visual cortex, the hippocampal distributions exhibit weaker modulation across stimulus conditions (Fig.~\ref{fig:kurts}B).

\begin{figure*}[ht!]
\centering
\includegraphics[width=\textwidth]{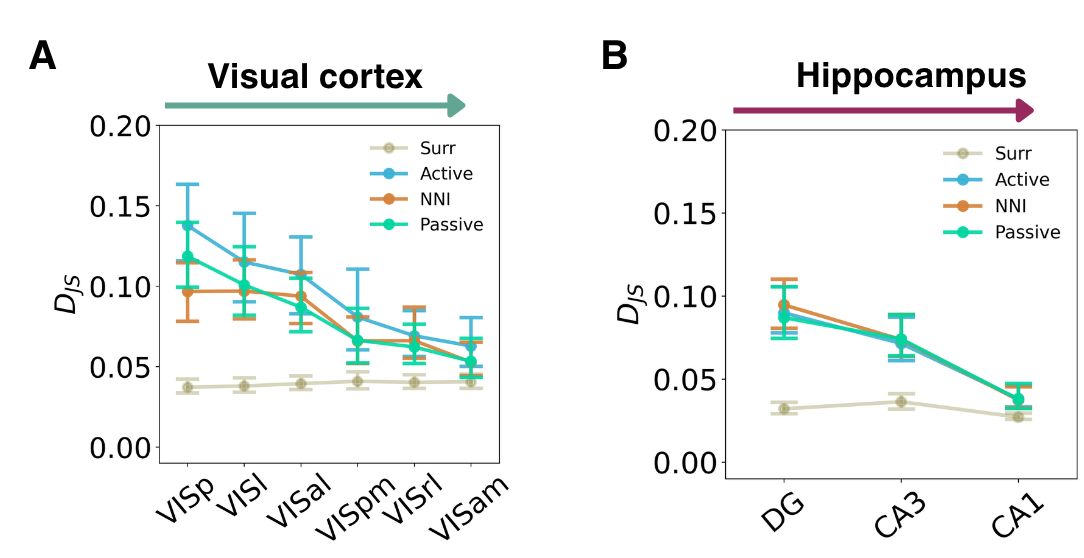}
\caption{Distances to criticality across brain regions.
(A) Median and quartiles of each $D_{JS}$ during all phases of the experiment. 
The distance  $D_{JS}$ decreases together with the number of connections to the primary visual cortex and is modulated by the experimental phase, increasing during the active phase compared to the other conditions.
(B) Median and quartiles of each $D_{JS}$ distribution for the hippocampus. The distances decay following the known hierarchical organization of this area. In this case there is no modulation with experimental phase.
}
\label{fig:dist}
\end{figure*}

To compare the strength of critical signatures across regions, we first examined the median $D_{JS}$ values. 
When regions are ordered according to the visual cortical hierarchy reported by Wang \textit{et al.}~\cite{wang_2012_newtorkanalysis}, which is based on the relative strength of projections from the primary visual area to downstream targets measured via optical density, the median $D_{JS}$ systematically decays (Fig.~\ref{fig:dist}A).
This decay is more pronounced during the active phase, whereas for the passive phase the decreasing sequence stays similar, but with  substantially smaller absolute values.
For non-natural images, the primary and lateral visual (VISp and VISl) regions present a plateau, while higher  visual areas present a profile very similar to that of the passive phase. 
Moreover, for the hippocampus the distances also decrease (Fig.~\ref{fig:dist}D) when regions are ordered according to their anatomical hierarchy~\cite{bienkowski2018integration}.

\subsection{A functional hierarchy of critical dynamics}

\begin{figure*}[ht!]
\centering
\includegraphics[width=\textwidth]{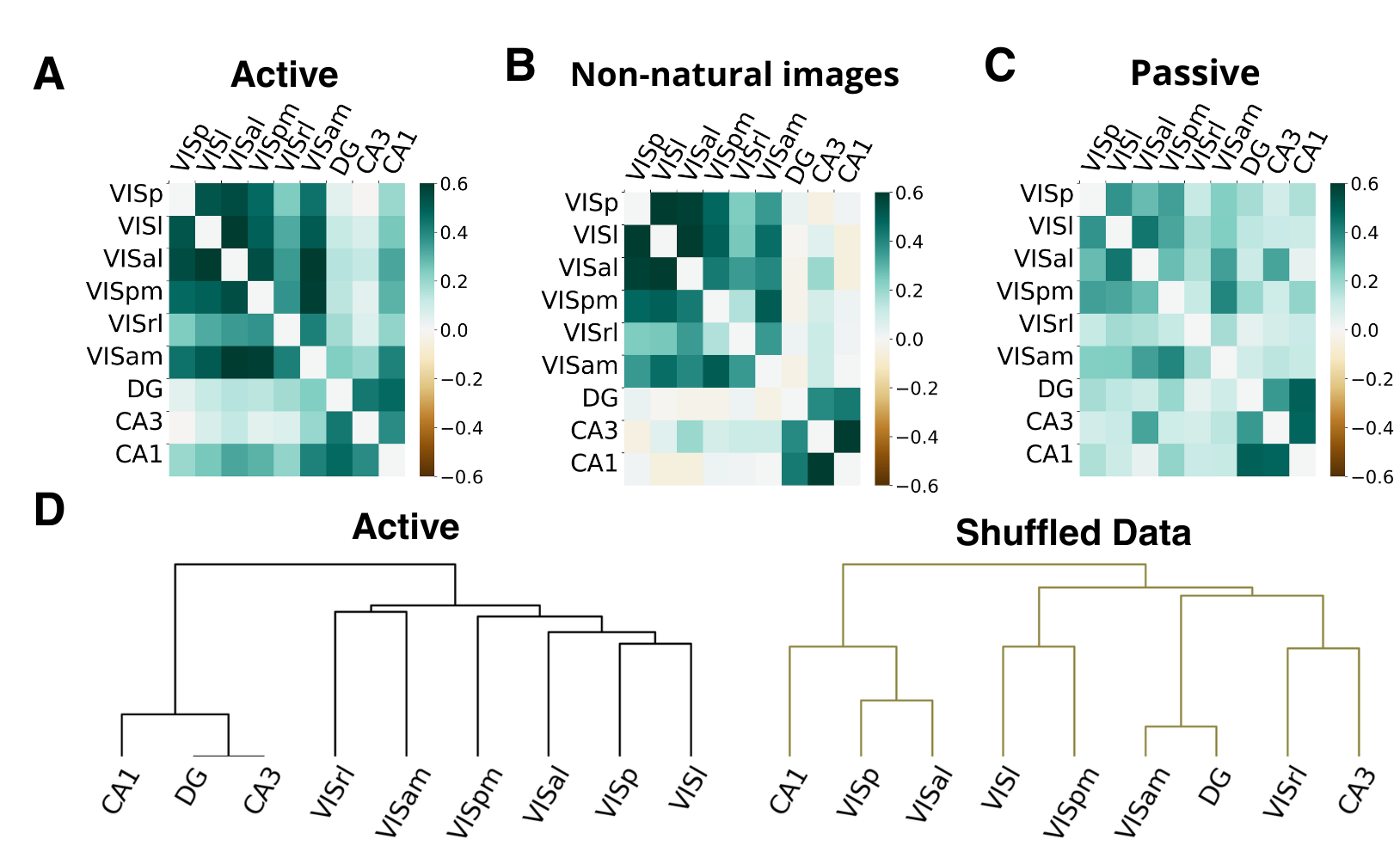}
\caption{Correlations of criticality signatures across different areas.
(A),(B) and (C) 
Correlations of $D_{JS}$ across different areas in the active, NNI, passive phases respectively. (D) Inferred hierarchy of critical dynamics. By applying a modularity algorithm to the correlations between $D_{JS}$ values for different areas during the active phase we are able to obtain a functional hierarchy of critical dynamics. The VISrl region, usually depicted as the base of the dorsal hierarchy, appears as the least functionally connected to the rest of the areas of the visual cortex available in this dataset.
}
\label{fig:corr}
\end{figure*}

We next investigated the functional coupling of critical dynamics between different regions. 
For each 30-second window of each brain region, we computed the $D_{JS}$ of the activity projected onto the K/32 eigenvectors. 
Next, we computed the Pearson correlation between the $D_{JS}$ time series for all pairs of regions, averaging across sessions in which both regions were available (see Methods).

The resulting correlation matrices (Fig.~\ref{fig:corr}) reveal a  functional network that depends on the phase of the experiment (active, passive, NNI).
A clear modular structure, with stronger coupling within the visual and hippocampal systems than between them, is visible in all the three phases.
The average correlation within the visual areas and within the hippocampus are respectively ${C_{vis}=0.5 \pm 0.1}$ and ${C_{hp}=0.43  \pm 0.04}$ for the active phase, ${C_{vis}=0.3 \pm 0.1}$ and ${C_{hp}=0.42 \pm 0.05}$ for the passive phase and ${C_{vis}=0.5 \pm 0.2}$ and ${C_{hp}=0.45\pm0.08}$ for non-natural images.

Crucially, the strength of cross-system correlation provides a direct estimation of the level of integration taking place between visual and hippocampal networks. This integration is strongest during phases involving natural images:
${C_{vis-hp}=0.2 \pm 0.1}$ for the active phase, ${C_{vis-hp}=0.13 \pm 0.06}$ for the passive phase and ${C_{vis-hp}=0.02 \pm 0.06}$ for non-natural images.

To extract the underlying hierarchy from the active task network, we performed a community detection analysis. 
We treated the correlation matrix as a weighted adjacency matrix, setting the diagonal to zero to ignore self-loops. 
We then applied a standard modularity optimization algorithm (see Methods) to obtain the multi-scale structure represented by the dendrogram of Fig.~\ref{fig:corr}D. 
It represents a functional hierarchy inferred purely from the critical dynamics.
This inferred hierarchy shows a remarkable correspondence to the known anatomical architecture of the mouse brain. 
At the broadest scale, the algorithm correctly partitions the hippocampal formation from the visual cortex. 
Within the visual system, it groups areas in a manner largely consistent with known processing streams~\cite{wang_2012_newtorkanalysis, bienkowski2018integration, dsouza2022hierarqui}.
The same consistency with anatomical results is also observed for the passive and NNI functional networks (see Supplementary Material). 
These correspondences are lost if the $D_{JS}$ time series are shuffled (Fig.~\ref{fig:corr}D). 
This result demonstrates that the correlation structure of criticality signatures contains a detailed blueprint of the brain's hierarchical organization.

To ensure the reliability of the extracted hierarchy, we detail several supporting analyses in the Supplementary Material. 
We confirm that this community structure is robust across the different experimental phases, and remains consistent when evaluated with an alternative criticality metric (the real-space PRG $\alpha$ exponent; see Methods).
Methodologically, we show that coarse-graining is an essential step for the $D_{JS}$ network to reveal this hierarchical structure. 
Crucially, applying our extraction procedure directly to population firing rates fails to produce a meaningful hierarchy during the active phase, indicating that this large-scale organization is uniquely captured by critical dynamics rather than simple rate fluctuations.


\subsection{Scaling Exponents}
 We can also employ a second type of renormalization, called real-space phenomenological renormalization~\cite{Meshulam2019coarse,nicoletti_scaling_2020,ponce-alvarez_critical_2023,castro_and_2024,munn_multiscale_org,cambrainha2025criticality,castro2025interdependent,nasciment2025arXiv,zivadinovic2026scaling}.
This method iteratively sums the activity of the most correlated neuron pairs, allowing for the calculation of different scaling exponents: $\alpha$, obtained from the mean variance of the activity across these coarse-graining steps; $\beta$, defined from the probability of finding silence within those clusters at each step; and $z$, which is extracted from the decay of the autocorrelation time  (see Methods). 
These exponents have known values for trivial behavior, namely: $\alpha=1$ (central limit theorem), $\beta=1$ (summed Poisson processes) and $z=0$ (characteristic time independent of system size).

\begin{figure*}[ht!]
\centering
\includegraphics[width=\textwidth]{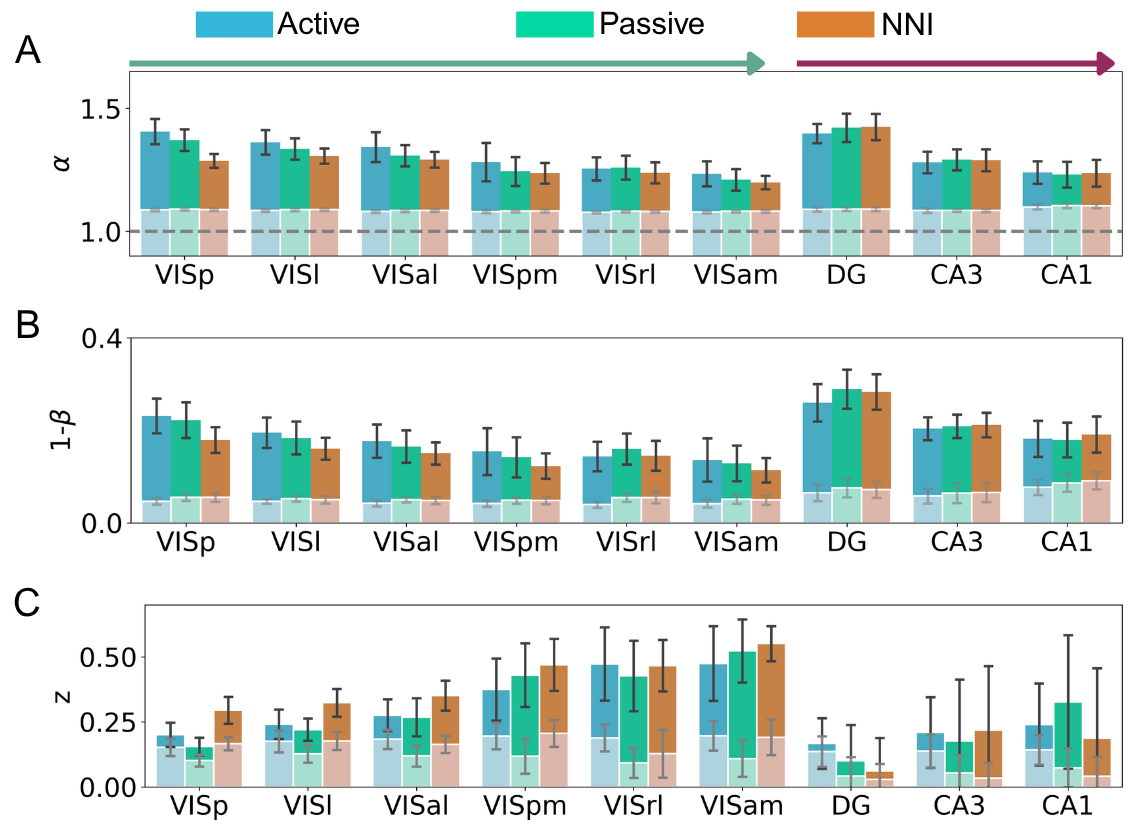}
\caption{Exponents from the real space phenomenological renormalization of each brain area during different experiment sections. (A) The exponent $\alpha$ varies across regions and experimental context, decaying along the morphological hierarchy in the visual cortex and the hippocampus. (B) The exponent $\beta$ follows a similar tendency to $\alpha$, showing lower values for higher hierarchical levels. (C) The exponent z shows the opposite trend, increasing with the hierarchical level. In the hippocampus the z exponents are closer to 0. Exponents from surrogate data are shown below their respective data in white colors.}
\label{fig:exp}
\end{figure*}

The approach revealed distinct organizational patterns across brain areas. 
In both visual cortex and hippocampus, $\alpha$ and $1-\beta$ decreased systematically along the known anatomical hierarchy (Figs.~\ref{fig:exp}A and~\ref{fig:exp}B). 
This tendency toward trivial values of real-space scaling exponents as one progresses through the anatomical hierarchy is consistent with the decrease in momentum-space criticality signatures (Fig.~\ref{fig:kurts}).
In contrast, the exponent $z$ exhibits the exact opposite trend with the highest value $z=0.55 \pm 0.07$ at VISam (Fig.~\ref{fig:exp}C). 
Together, the scaling exponents suggest a non-trivial structure, with the exponents $\alpha$ and $\beta$ reaching values near $\alpha = 1.42 \pm 0.05$ and $\beta = 0.72 \pm 0.04$ in the DG (Fig.~\ref{fig:exp}A). 

To statistically validate the observed hierarchical gradients, we compared the scaling exponents ($\alpha$, $\beta$, $z$) across brain regions using a one-way ANOVA, conducted independently for the visual cortex and hippocampus during each experimental phase. 
We then performed pairwise comparisons between regions, adjusting all $p$-values with the Bonferroni correction. 
Complete statistical tables are provided in the Supplementary Material.

\begin{figure}[ht!]
\centering
\includegraphics[width=0.49\textwidth]{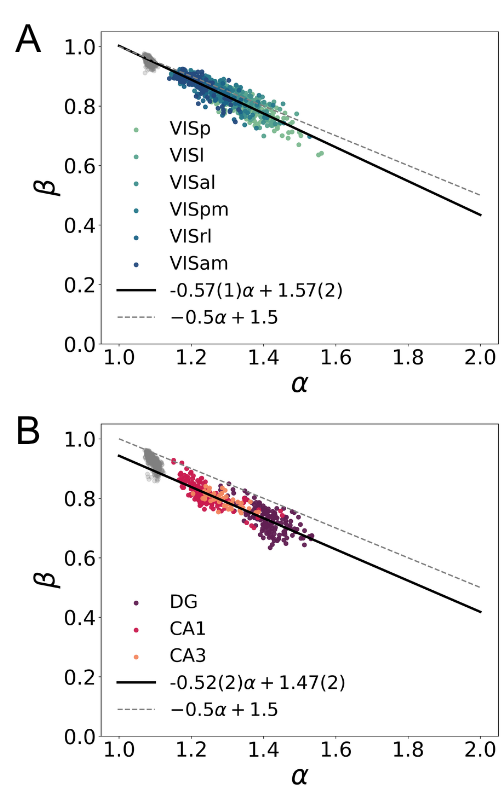}
\caption{Exponents relation across brain areas. (A) Relationship between PRG exponents $\alpha$ and $\beta$ for visual cortical regions. Gray points represent surrogate controls. Visual regions lie close to the theoretical prediction $\beta = \frac{(3-\alpha)}{2}$, and the empirical fit closely matches the predicted slope. (B) Same analysis as in (A), for hippocampus. The points appear shifted away from the theoretical line, but their fit remains nearly parallel to the prediction and the global offset is small. The corresponding nonlinear relationships between these exponents and the exponent $z$ are presented in Fig.~S6 of the Supplementary Material.}
\label{fig:exp_rel}
\end{figure}

We can also verify that the obtained exponents closely follow a recently predicted theoretical relation~\cite{castro2025interdependent}:
\begin{equation}
\beta = \frac{3 - \alpha}{2}.
\end{equation}
This agreement holds across brain regions and experimental conditions (Fig.~\ref{fig:exp_rel}), providing empirical support for the theoretically predicted interdependence between the exponents~\cite{castro2025interdependent}. Notably, anatomically and functionally distinct regions all collapse onto the same scaling relation, suggesting that this law applies broadly across neural systems despite substantial differences in local organization.
Furthermore, while there are currently no theoretical predictions relating the exponents $\alpha$ and $\beta$ to the dynamical exponent $z$, our data suggest a nonlinear relationship between them (see Supplementary Material for detailed plots).

\section{Discussion}

Building upon foundational works that, for decades, have supported the idea that the brain operates near a critical point~\cite{linkenkaer-hansen_long-range_2001,beggs_neuronal_2003, hengen_shew_is_criticality}, our research addresses the subsequent question of how the signatures of criticality are distributed across the brain's architecture.
We have shown that these signatures are not uniform, but instead exhibit a clear hierarchical organization in both the mouse visual cortex and hippocampus. 
This finding highlights the importance of the structural hierarchy in sculpting spatiotemporal critical dynamics. 
However, a notable dissociation emerges between markers of criticality associated with static versus dynamical features. 
While the exponents $\alpha$ and $\beta$, more closely related to static properties~\cite{munn_multiscale_org}, suggest that peripheral regions at the lower levels of the hierarchy lie closer to criticality, the dynamical exponent $z$ instead indicates that hub regions at the top of the hierarchy are closer to criticality.
Furthermore, we have shown that the anatomical hierarchy can be accurately reconstructed from the temporal correlations between criticality signatures.
In the visual cortex, but not in the hippocampus, this organization is dynamically modulated by context, with task engagement altering the landscape of criticality in the visual system. 

On the one hand, this ``hierarchy of criticality'' observed from $\alpha$ and $1-\beta$ suggests that cortical processing is structured by graded proximity to critical dynamics.
For this dataset, recorded during a visual task, the primary visual cortex exhibits the strongest critical signatures. 
Being the main recipient of complex visual information among the areas in this study, it is reasonable to expect it to be maintained at a state of maximal sensitivity to input~\cite{kinouchi_optimal_2006}.
As information propagates to higher-order areas for more specialized processing, the system may shift its dynamical regime away from this point of high susceptibility. 
In contrast, the hippocampal signatures of criticality are far less sensitive to context, suggesting that its dynamics are governed more by intrinsic computational demands, such as memory processes.

On the other hand, the opposite trend was observed for the exponent $z$, which captures how the autocorrelation characteristic time scales with cluster size. From this perspective, regions higher in the hierarchy have more non-trivial properties associated with closer distance to criticality. As dynamical systems slowly approach a critical point, the temporal correlations are stretched, which is known as critical slowing down. This argument has been previously explored to establish a link between criticality and the hierarchy of the visual cortex of the mouse~\cite{harris2024tracking}. Furthermore, this can be considered a neural basis to support a hierarchy of timescales in the brain~\cite{kiebel2008hierarchy,murray2014hierarchy, gollo2015dwelling}. In this sense regions that are higher in the hierarchy will exhibit fluctuations following longer temporal scales.

These seemingly contradictory findings regarding the direction of the hierarchical gradient underscore the central role of anatomical organization in constraining brain dynamics. While studies based on invasive electroencephalography (iEEG) recordings have reported a consistent relationship between spatial and temporal correlations~\cite{muller2023spatial}, in line with expectations from critical phenomena, more recent work points to a more nuanced picture. For instance, fMRI studies have revealed opposing gradients, with higher-order regions exhibiting longer temporal correlations but lower signal variance~\cite{Ponce2025network}, challenging established principles from criticality theory. Complementing this view, another work~\cite{harris2024tracking}, also based on the Neuropixels recordings from the Allen Institute, has shown that canonical indicators of distance to criticality, such as standard deviation and autocorrelation, are reliable only under conditions of fixed noise levels. In more realistic settings with variable noise, as expected in behaving animals, these measures lose sensitivity, whereas alternative dynamical indicators provide more robust estimates. These approaches reveal a gradient in which higher-order regions lie closer to criticality~\cite{harris2024tracking}. Together, this fast-growing body of work suggests that static and dynamical properties capture distinct aspects of brain organization and may play different roles in shaping collective dynamics~\cite{Ribeiro10a,cambrainha2025criticality,Ponce2025network}. It also highlights that a comprehensive understanding of brain criticality remains an open and evolving challenge.

Notably, the level of task engagement not only modulates the criticality signatures within individual regions, as previously reported~\cite{cambrainha2025criticality}, but also modulates the correlation between the signatures of criticality in different areas.
During the active phase, mice are exposed to the exact same images as during the passive phase, but only in the former can they actively engage in the task and receive reward. 
This difference is reflected in the corresponding functional networks, which in the active phase show much higher integration, along with stronger cortical-hippocampal connectivity. 
While it may be premature to speculate on the exact mechanism underlying these observations, it hints at a link between behavior --- in this case, possibly attention --- and non-trivial neural dynamics across different regions exhibiting greater integration.

The relationship between structural and functional connectivity in neuroscience is notoriously known to be a difficult problem~\cite{honey2009predicting, fotiadis2024structure}.
In the Allen Visual Behavior Neuropixels dataset, this is no different: the functional connectivity matrix obtained from firing rates is a poor predictor of the underlying structural connectivity. 
Surprisingly, however, this novel form of functional connectivity explored here obtained from signatures of criticality (Jensen-Shannon distances between PRG-renormalized activities, or PRG exponents) reveals fundamental organizational principles grounded in the network hierarchy that stems from the structural connectivity. This finding opens promising avenues for applying this approach in other experimental paradigms.

A notable exception to the consistence of the inferred hierarchy from the functional network is the rostrolateral visual area (VISrl). In our functional network, it separates from the rest of the visual cortex on the first few steps, though it is usually reported as being structurally closer to the primary visual cortex~\cite{dsouza2022hierarqui}. 
Interestingly, the anomalous clustering of the VISrl points to a unique functional role for this region, as  recently reported by D'Souza \textit{et al.}~\cite{dsouza2022hierarqui}.

Since our findings are correlational, further studies may investigate to which extent a specific network architecture is a necessary condition for the existence of the ``criticality hierarchy'' we observe here. It also remains to be determined whether the difference in the $D_{JS}$ profile between the active and passive phases (Fig.~\ref{fig:dist}A) are due to the learning of the specific task in this experimental setup or naturally modulated by endogenous activity. Finally, one can ask whether each brain area is locally maintained at a distinct distance from criticality, or if a single region near criticality propagates its statistical signatures through network coupling~\cite{rabuffo2025connectome}.

In the hippocampus, particularly in the Dentate Gyrus, some time windows exhibit activity distributions with a broader central peak and lighter tails (see Supplementary Material) compared to the more typical highly non-Gaussian windows (Fig.~\ref{fig:kurts}). 
This shift in the shape of the distribution of renormalized activity points to a potentially different nature of the phase transition driving the dynamics in these regions.  
When analyzed through the lens of real-space renormalization, the resulting exponents are also highly non-trivial (Fig.~\ref{fig:exp}). 
This agreement between both methods confirms that the underlying critical dynamics are robust.

We also observe that the scaling exponents closely follow a predicted theoretical relation, regardless of brain region or experimental phase. 
This theoretical relation was originally derived for thresholded human fMRI data by assuming that coarse-grained activity follows an exponential distribution and that the silence threshold exhibits a non-linear scaling with system size~\cite{castro2025interdependent}. 
Finding the same agreement in the high-resolution spiking activity of mice indicates that this theoretical relation hold much more broadly.

Our results highlight that critical dynamics in the brain are not a monolithic phenomenon but possess a rich spatial organization that mirrors its morphological hierarchy. 
By showing that this structure can be read out directly from the dynamics, we highlight a connection between the physics of collective activity to the principles of systems neuroscience, providing a quantitative foundation for understanding how the brain's architecture and its emergent dynamics are fundamentally intertwined to give rise to function.

\begin{acknowledgments}

The authors acknowledge support from Brazilian agencies CNPq (Grants No.~140660/2022-4, No.~308703/2022-7, No.~314094/2023-7, No.~444500/2024-3 and No.~408389/2024-9), FACEPE (Grant No.~APQ-1187-1.05/22) and FADE/UFPE (Grant No.~64/2024). LLG was supported by the Ramón y Cajal Fellowship (RYC2022-035106-I); the project ANCHOR (PID2024-162343OB-I00), funded by MICIU/AEI/10.13039/501100011033/FEDER, UE; and the María de Maeztu Program for Units of Excellence (CEX2021-001164-M), funded by MICIU/AEI/10.13039/501100011033.

\end{acknowledgments}

\appendix

\section{Methods and Materials}\label{materiaslandmethods}
\subsection{Data}\label{secdata}
For this study, we used a publicly available dataset provided by the Allen Brain Institute~\cite{allendandi}, which consists of large-scale neuronal recordings from multiple brain areas in mice using Neuropixels 1.0 probes. 
Each recording session lasted approximately 2 hours and 25 minutes.

The experiment was divided into three main parts. 
In the first part, mice performed a visual recognition task, where they were rewarded for correctly identifying an image change by licking a spout. 
The second part consisted of a 25-minute passive presentation of a sequence of Gabor patches, black and white flashes, and a gray screen. 
In the third part, the same sequence of natural images from the first part was shown again, but under passive viewing conditions without a task or reward (lick port was retracted).

Our analysis focused on a subset of these data. We first identified all brain regions within the visual cortex and hippocampus that had at least 15 recording sessions meeting our criteria.
From this pool of regions, listed in Table~\ref{tabela}, we selected all sessions containing more than 128 recorded units (including both single-unit and multi-unit activity). 
These thresholds were chosen to ensure robust application of our analytical methods (128-unit minimum) and statistical robustness (15-session minimum).

Finally, the spike trains from each selected session were binned into 50-ms intervals, binarized and segmented into 30-s windows.
The first window from sessions 1118324999 and 1118512505 was discarded due to artifacts caused by a misalignment between the logged session start-time and the onset of recorded neural activity.

\begin{table*}[]
    \centering
    \begin{tabular}{|c|c|c|c|}
    \hline

         \textbf{Brain region} &  \textbf{Acronym} & \textbf{Analyzed sessions} & \textbf{Unit count per session} \\
         & & & \textbf{(mean $\pm$ standard deviation)} \\
             \hline
                Primary visual    & VISp & 62 & $165\pm27$ \\
                    \hline
                Lateral visual    & VISl & 48 & $163 \pm 32$\\
                    \hline
                Anterolateral visual    & VISal & $32$ & $155 \pm 30$\\
                    \hline
                Posteromedial visual     & VISpm & 57 & $156\pm 24$\\
                    \hline
                Rostrolateral visual    & VISrl & 26 &$149 \pm 18$\\
                    \hline
                Anteromedial visual    & VISam & 31 & $151 \pm 22$\\
                    \hline
                Dentate gyrus & DG & 70 & $202 \pm 61$\\
                    \hline
                Cornu ammonis 3 & CA3 & 18 & $171 \pm 41$\\
                    \hline
                Cornu ammonis 1 & CA1 & 57 & $329 \pm 109$\\
                    \hline
                
    \end{tabular}
    \caption{Brain regions considered in this study, their respective acronym, number of experimental sessions analyzed and units counts (mean $\pm$ standard deviation) per session. Both single-unit and multi-unit activity (SUA/MUA) are included.}
    \label{tabela}
\end{table*}

\subsection{Surrogate data}
To evaluate how our results differ from chance, we generated surrogate data by shuffling the inter-spike intervals of each unit within the 30-s time window. This procedure eliminates correlations between spikes while preserving firing statistics, providing reference values for comparison.

\subsection{Momentum-space phenomenological renormalization}

To coarse-grain the activity and obtain normalized distributions, we employ the phenomenological renormalization procedure~\cite{bradde_pca_2017,nicoletti_scaling_2020,castro_and_2024}. 
Starting from the covariance matrix,
\begin{equation}\label{cov}
    C_{ij} = \langle \sigma_i \sigma_j \rangle,
\end{equation}
where $\sigma_i(t)$ denotes the time series of activity of unit $i$, we compute its ranked eigenvalues ${\lambda_1 > \lambda_2 > \ldots > \lambda_N}$ and corresponding eigenvectors $u_i$. 
We then construct projectors into subspaces of binary 50-ms binned activity dimension $N_\mathrm{cutoff}$ as
\begin{equation}\label{equ3}
    \hat{P}_{ij}(N_\mathrm{cutoff}) = \sum_{\mu=1}^{N_\mathrm{cutoff}} u_{\mu i} u_{\mu j},
\end{equation}
from which we define a new set of renormalized variables:
\begin{equation}
    \psi_i(N_{\mathrm{cutoff}}) = \mathcal{Z}_i(N_{\mathrm{cutoff}}) 
    \sum_{j=1}^{N} \hat{P}_{ij}(N_{\mathrm{cutoff}}) \left( \sigma_j - \langle \sigma_j \rangle \right).
\end{equation}
This procedure effectively coarse-grains the system by progressively removing eigenmodes associated with smaller eigenvalues, thereby filtering finer scales of activity. 
We apply it for $N_{\mathrm{cutoff}} = N,\, N/2,\, N/4,\, N/8,\, N/16,\, N/32$.

We calculate the Jensen-Shannon divergence \cite{lin1991divergence} between the distribution of the most coarse-grained normalized activity ($N_{\mathrm{cutoff}} = N/32$), defined as
\begin{equation}
P(\psi) = \frac{1}{N}\sum_{i=1}^N \mathbb{P}[\psi_i(N_{\mathrm{cutoff}})=\psi],
\label{MSdist}
\end{equation}
and a standard Gaussian reference ($\mathcal{N}$) with zero mean and unit variance. 
The divergence is given by:
\begin{widetext}
\begin{equation}
    JS_{\mathrm{Div}}(P(\psi)||\mathcal{N}) = \frac{1}{2}\int{P(\psi)\log_2\left(\frac{P(\psi)}{M(\psi)}\right)d\psi}+\frac{1}{2}\int{\mathcal{N}(\psi)\log_2\left(\frac{\mathcal{N}(\psi)}{M(\psi)}\right)d\psi},
\end{equation}
\end{widetext}
with
\begin{equation}
    M=\frac{P+\mathcal{N}}{2}.
\end{equation}
From this, the Jensen-Shannon distance is given by:
\begin{equation}
    D_{JS}=\sqrt{JS_{\mathrm{Div}}(P(\psi)||\mathcal{N})}
\end{equation}
This metric is symmetric, satisfies the triangle inequality, and is zero only if $P=\mathcal{N}$ \cite{endres2003anewmetric}. 
Using a base-2 logarithm bounds this distance between 0 and 1. 
As the system is coarse-grained, non-critical dynamics converge toward a Gaussian ($D_{JS} \to 0$), whereas critical dynamics retain a non-zero distance, reflecting scale-invariant behavior.

\subsection{Jensen-Shannon distance correlations}
Applying the PRG procedure within 30-s windows yields a time series of $D_{JS}$ values for each experiment. 
To investigate the dynamics of brain activity, we computed the Pearson correlation between the $D_{JS}$ time series of two different brain areas within the same session:
\begin{equation}
    \mathrm{Corr}(D_{JS}^i(t), D_{JS}^j(t)) = \frac{\mathrm{Cov}(D_{JS}^i, D_{JS}^j)}{\sigma_{D_{JS}^i} \sigma_{D_{JS}^j}},
\end{equation}
where $D_{JS}^i(t)$ and $D_{JS}^j(t)$ denote the $D_{JS}$ time series of areas $i$ and $j$, respectively, and $\sigma_{D_{JS}}$ their standard deviations. 
We then averaged the correlations across all sessions in which the same pair of brain areas was recorded, constructing the final correlation matrix shown in Fig.~\ref{fig:corr}.

\subsection{Modularity and communities}
We employed the \texttt{networkx} Python library~\cite{networkx} to compute community structure in the network using the Louvain algorithm. The quality of a given partition was assessed using the modularity function~\cite{newman2004fiding,reichardt2006statistical}, defined as
\begin{equation}
    Q = \frac{1}{2M} \sum_{i \ne j} \left( A_{ij} - \gamma \frac{k_i k_j}{2M} \right) \delta(c_i, c_j),
\end{equation}
where $A_{ij}$ denotes the adjacency matrix, $k_i$ and $k_j$ are the degrees of nodes $i$ and $j$, $M$ is the total number of edges, $\gamma$ is the resolution parameter, and $\delta(c_i, c_j)$ equals 1 if nodes $i$ and $j$ are assigned to the same community and 0 otherwise. In our analysis, the adjacency matrix was constructed from the correlations between $D_{JS}$ values, providing a functional connectivity representation of criticality.

The modularity $Q$ quantifies the density of edges inside communities compared to a random null model, with higher values indicating stronger community structure. To explore the effect of resolution, we systematically varied $\gamma$ and recorded the resulting changes in community partitions, which are summarized in Fig.~\ref{fig:corr}D.

\subsection{Phenomenological renormalization exponents}

To extract PRG scaling exponents shown in Fig.~\ref{fig:exp} and Fig.~\ref{fig:exp_rel}, we employ an alternative phenomenological renormalization scheme based on real-space coarse-graining~\cite{Meshulam2019coarse}. The procedure proceeds by iteratively merging neural activity variables according to their pairwise correlations. At each step, the two neurons with the strongest correlation are grouped together, and their activities are summed to form a new coarse-grained variable,
\begin{equation}
\sigma_i'^{(k+1)} = \sigma_i^{(k)} + \sigma_j^{(k)},
\end{equation}
where neuron $j$ denotes the most strongly correlated partner of neuron $i$. Repeating this process $k$ times yields a reduced description consisting of $N/2^k$ coarse-grained units (clusters), each representing the combined activity of $K=2^k$ original neurons.

We track several observables that are expected to exhibit nontrivial power-law behavior only in the vicinity of a critical point. For each animal, these quantities are first averaged across time windows at each level of coarse-graining. Scaling exponents are then obtained by fitting the dependence of these averages on the cluster size $K$.

\paragraph{\normalfont \textit{Mean variance}}
We compute the variance of the activity within each cluster and denote by $M_2$ the variance averaged across all clusters of size $K$. This quantity scales with cluster size as
\begin{equation}
M_2 \propto K^\alpha.
\end{equation}
In the absence of correlations, the central limit theorem predicts the trivial scaling $\alpha=1$, whereas perfectly correlated activity yields $\alpha=2$.

\paragraph{\normalfont \textit{Silence probability}}
We define an effective free energy using the probability $P_{\mathrm{silence}}$ that a cluster exhibits no activity within a time bin~\cite{meshulam2018coarse}. Specifically,
\begin{equation}
F = -\log(P_{\mathrm{silence}}) \propto K^\beta.
\end{equation}
For independent Poisson spiking, this construction leads to the trivial exponent $\beta=1$.

\paragraph{\normalfont \textit{Temporal correlations}}
For clusters of size $K$, we compute the average autocorrelation function $C(t)$. From this, we extract a characteristic correlation time $\tau_c$, defined as the time lag at which the autocorrelation decays below $C(\tau_c)=0.1$. The dependence of this timescale on cluster size follows
\begin{equation}
\tau_c \propto K^z.
\end{equation}
For uncorrelated spike trains, the expected scaling is ${z=0}$.

\bibliography{ggcad}

\pagebreak
\clearpage
\begin{widetext}

\setcounter{section}{0}
\renewcommand{\thesection}{\arabic{section}}

\begin{center}
	\Large \textbf{Supplementary Material}
\end{center}
\vspace{1em}

\setcounter{figure}{0}
\renewcommand{\thefigure}{S\arabic{figure}}

\setcounter{table}{0}
\renewcommand{\thetable}{S\arabic{table}}

\setcounter{equation}{0}
\renewcommand{\theequation}{S\arabic{equation}}

	\section*{Hippocampus can present different shape of activity distribution}

	As discussed in the main text, the distribution of coarse-grained activity in the hippocampus typically exhibits a leptokurtic profile, characterized by a sharp peak and heavy tails. 
	However, certain time windows present a distinctly platykurtic shape. 
	In these instances, the distribution is broader than a standard Gaussian, yielding kurtosis values that are lower than those observed in the corresponding surrogate data. 
	Figure~\ref{fig:plato} illustrates one such window from the Dentate Gyrus during the passive phase, highlighting this broader structure in both logarithmic and linear scales.

	\begin{figure*}[ht!]
		\centering
		\includegraphics[width=\textwidth]{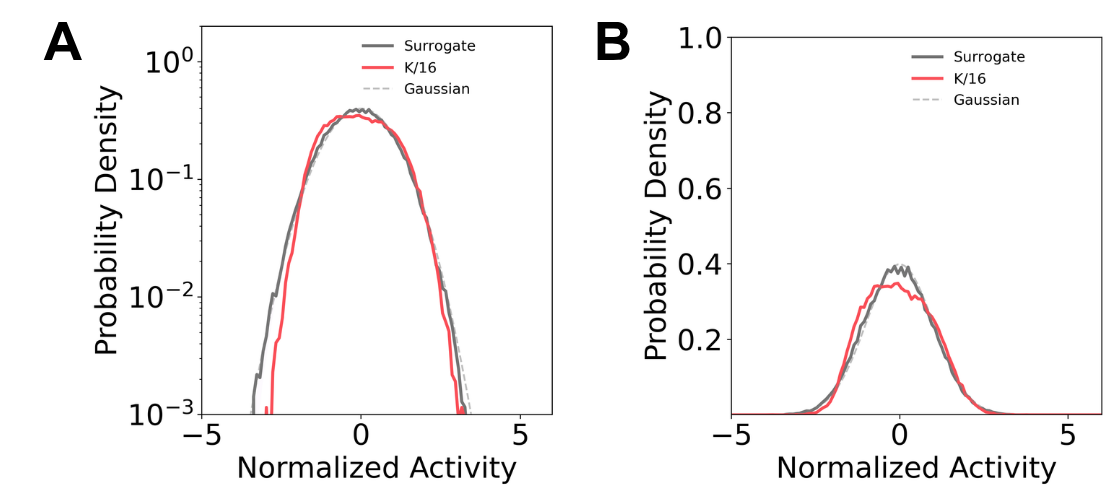}
		\caption{\textbf{Coarse-grained activity distribution for a DG window}. (A) and (B) display the distribution of normalized coarse-grained activity of the same time window. Though the surrogate follows closely the dotted gaussian line, the data shows a plateau around the zero.}
		\label{fig:plato}
	\end{figure*}

	\section*{Community structure for the other experimental phases}
	
	Applying the community detection procedure described in the main text to the passive and non-natural image (NNI) phases yields functional hierarchies that are highly consistent with the active phase. 
	As shown in Figure~\ref{fig:comum_other}, the multiscale organization remains stable across different experimental conditions, preserving the core structural relationships of the visual and hippocampal systems.
	
	\begin{figure*}[ht!]
		\centering
		\includegraphics[width=\textwidth]{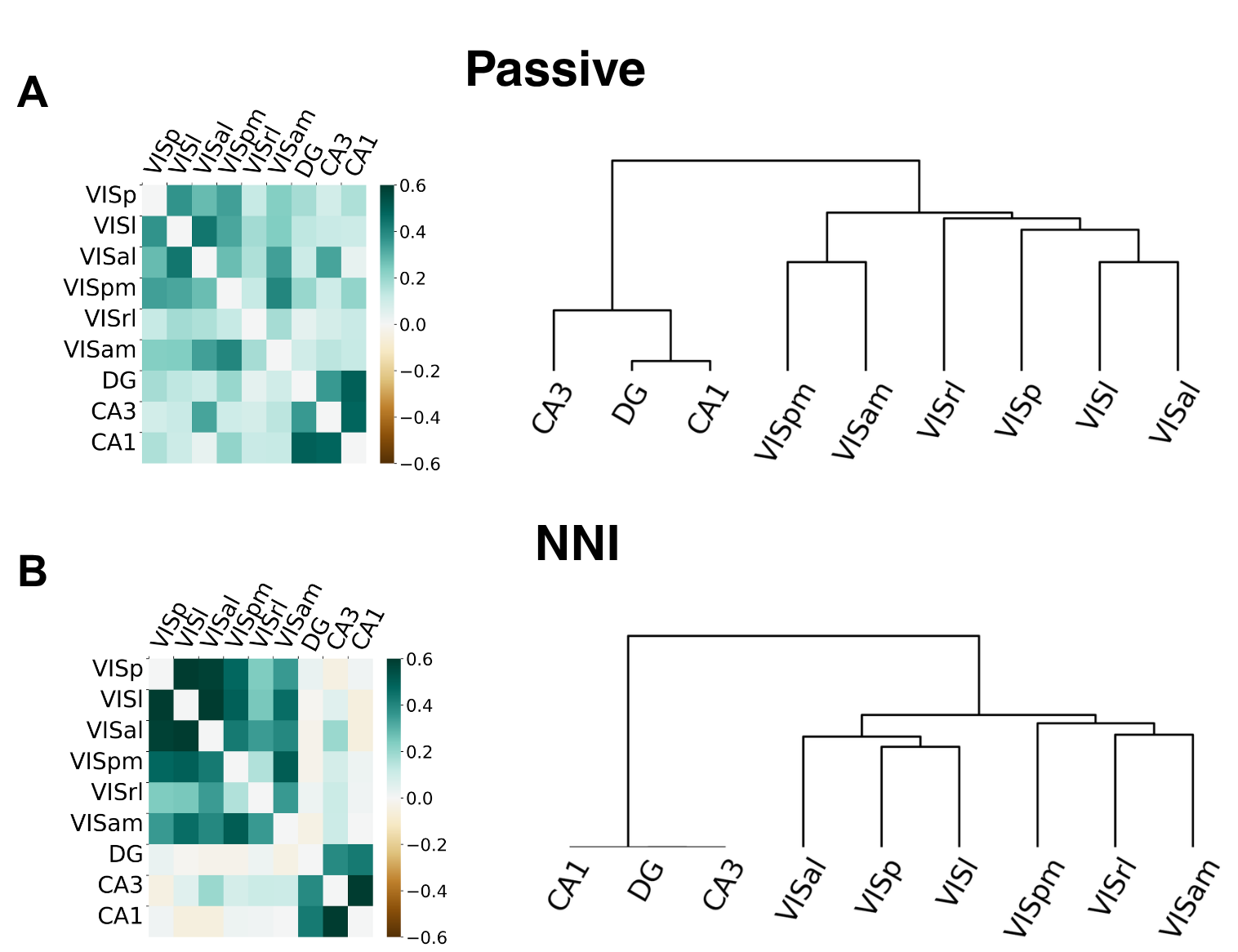}
		\caption{\textbf{Community structure during different experimental phases.} (A) and (B) the resulting community structure of both the passive and the NNI phases of the experiment display similar behavior that what is observed during the active phase, that is, it is similar to what is observed in the structural connections.}
		\label{fig:comum_other}
	\end{figure*}

	\section*{Community structure using population activity}
	
	To verify that the hierarchical organization is a specific feature of critical dynamics rather than a generic property of the firing rates, we examined the correlation matrices derived directly from the population activity. While population activity provides a simpler metric that could theoretically encode community structure, applying our clustering procedure to these correlations during the active phase fails to recover the known anatomical hierarchy. As shown in Figure~\ref{fig:commun_pop}, the resulting dendrogram mixes hippocampal and visual areas without reflecting the underlying structural organization.
	\begin{figure*}[ht!]
		\centering
		\includegraphics[width=\textwidth]{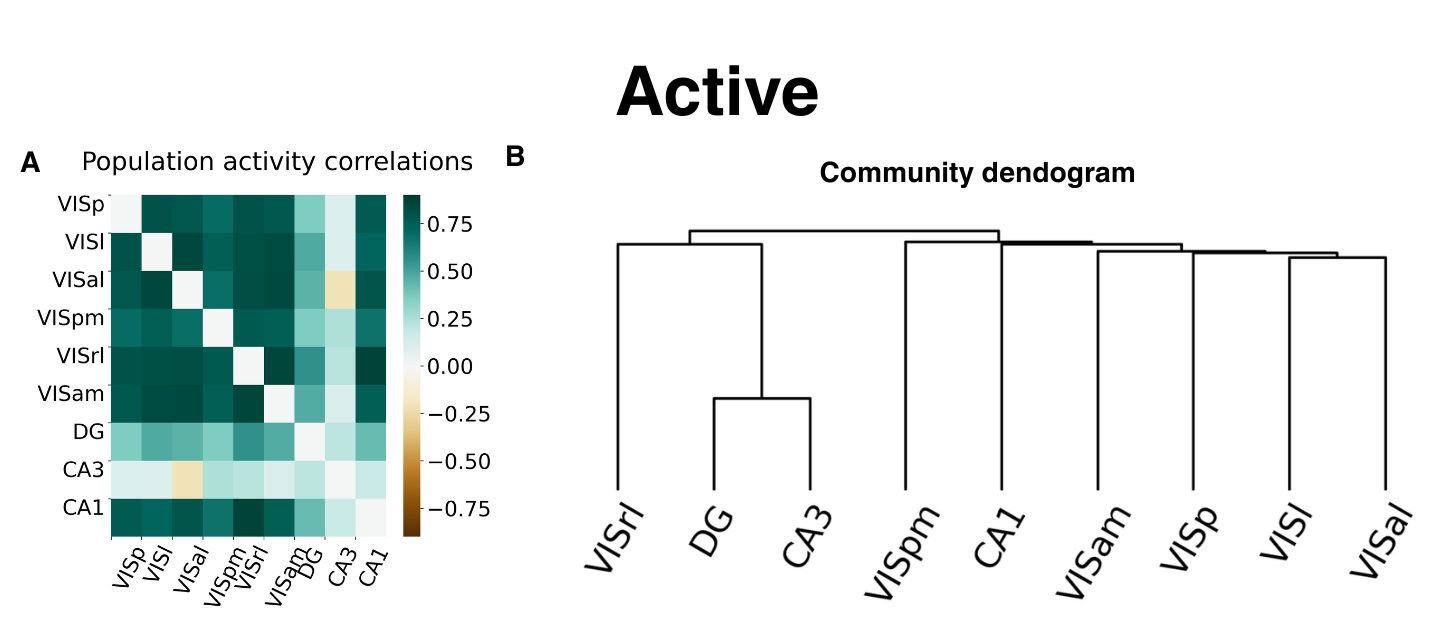}
		\caption{\textbf{Community structure of population activity.} (A) The correlations obtained from the population activity itself are much higher than what we observe from the kurtosis of the coarse-grained distribution. (B) The resulting community structure mixes hippocampus and visual areas and show no clear relation with the structural organization of the brain.}
		\label{fig:commun_pop}
	\end{figure*}
	
	\section*{Community structure using critical scaling exponent alpha}
	
	As an alternative to the momentum-space $D_{JS}$ metric, we can also reconstruct the functional hierarchy using the real-space scaling exponent $\alpha$, which serves as an independent signature of criticality. 
	By computing $\alpha$ for each 30-second window and calculating the pairwise correlation of these time series across brain regions, we obtain a functional network based purely on spatial scaling.
	The resulting community structure (Fig.~\ref{fig:alphacommun}) closely mirrors the one obtained with $D_{JS}$, confirming that the hierarchical organization of critical dynamics is robust to the choice of signature.
	\begin{figure*}[ht!]
		\centering
		\includegraphics[width=\textwidth]{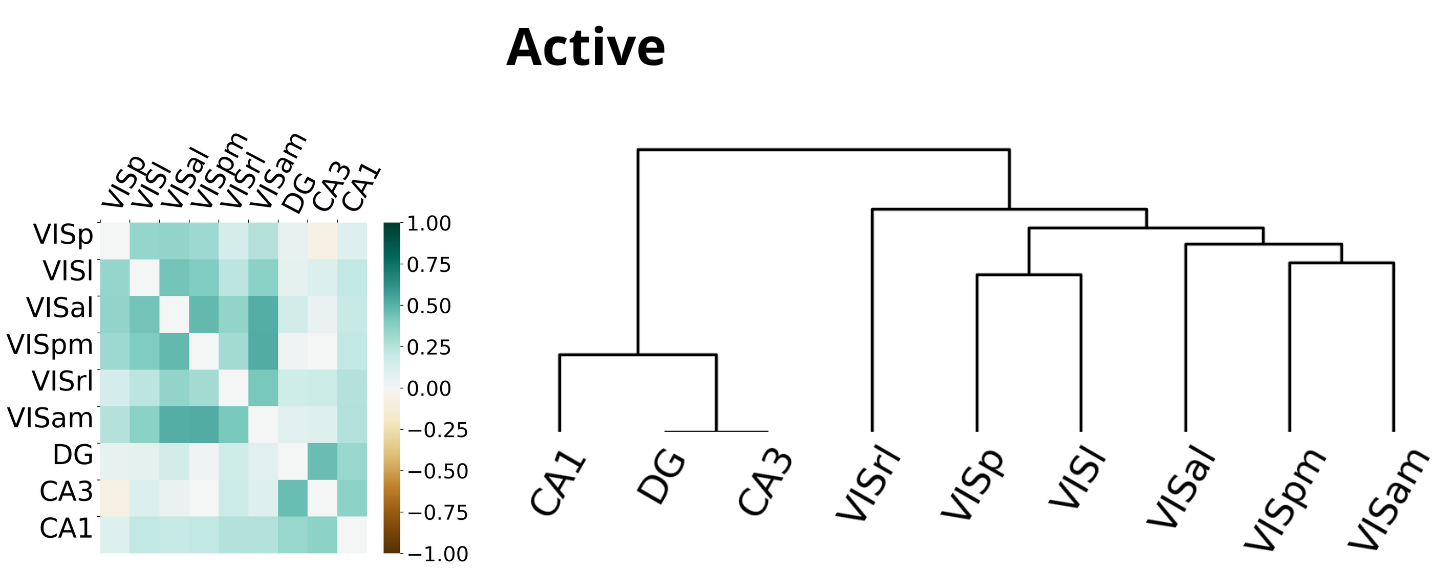}
		\caption{\textbf{Community structure derived from the real-space scaling exponent $\alpha$.} The dendrogram represents the hierarchical clustering of brain regions based on the temporal correlations of the exponent $\alpha$ during the active experimental phase. The resulting community structure is highly consistent with the network obtained using $D_{JS}$, accurately reflecting the known anatomical connections.}
		\label{fig:alphacommun}
	\end{figure*}

	\section*{Community detection for the Jensen-Shannon distance without coarse-graining}
	Although population firing rates do not reflect the anatomical hierarchy, it is possible that the full distribution of raw activity might. 
	We tested this by computing the Jensen-Shannon distance to a Gaussian baseline without applying any coarse-graining step. 
	As shown in Fig.~\ref{fig:nocoarse}, this method fails to recover the structural organization. 
	The resulting community structure arbitrarily aggregates different brain regions without reflecting their structural connectivity, closely mirroring the failure of the population activity control.
	
	\begin{figure*}[ht!]
		\centering
		\includegraphics[width=\textwidth]{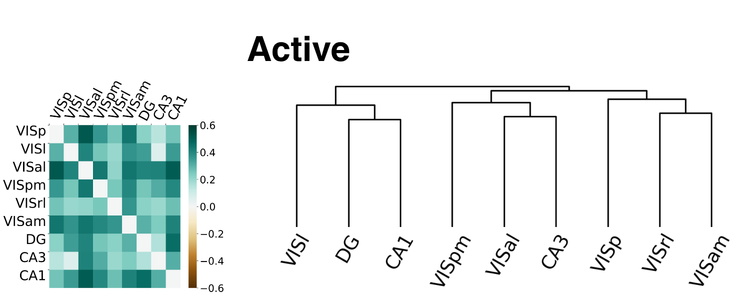}
		\caption{\textbf{Community structure of normalized activity without coarse-graining.} (A) The correlations obtained from
			normalized activity without coarse-graining. (B) The resulting community structure mixes hippocampus and visual areas
			and show no clear relation with the structural organization of the brain.}
		\label{fig:nocoarse}
	\end{figure*}
	
	\section*{Other metrics of non-gaussianity}
	In the main text, we quantified deviations from Gaussianity using the Jensen-Shannon distance ($D_{JS}$). This metric was selected because it acts as a true, symmetric mathematical distance. However, our results are robust to the specific choice of metric. Computing the distance between the coarse-grained normalized activity and a standard Gaussian using alternative measures such as 
	the Kullback-Leibler divergence~\ref{fig:kl_dists} and the Wasserstein distance~\ref{fig:wass_dists} yields similar results.
	
	\section*{Scaling relation between $\alpha$, $\beta$ and $z$}
	While the main text demonstrates that the static exponents $\alpha$ and $\beta$ closely follow a theoretical scaling relation~\cite{castro2025interdependent}, their relationship with the dynamical scaling exponent $z$ lacks a formal theoretical derivation. 
	However, the empirical data reveals a clear coupling between the static and dynamical exponents, particularly within the visual cortex (Fig.~\ref{fig:alpha_z}). As static signatures of criticality decay deeper into the hierarchy, the dynamical signatures increase, resulting in a distinct inverse relationship between $z$ and both $\alpha$ and $1-\beta$.

	\begin{figure*}[ht!]
		\centering
		\includegraphics[width=\textwidth]{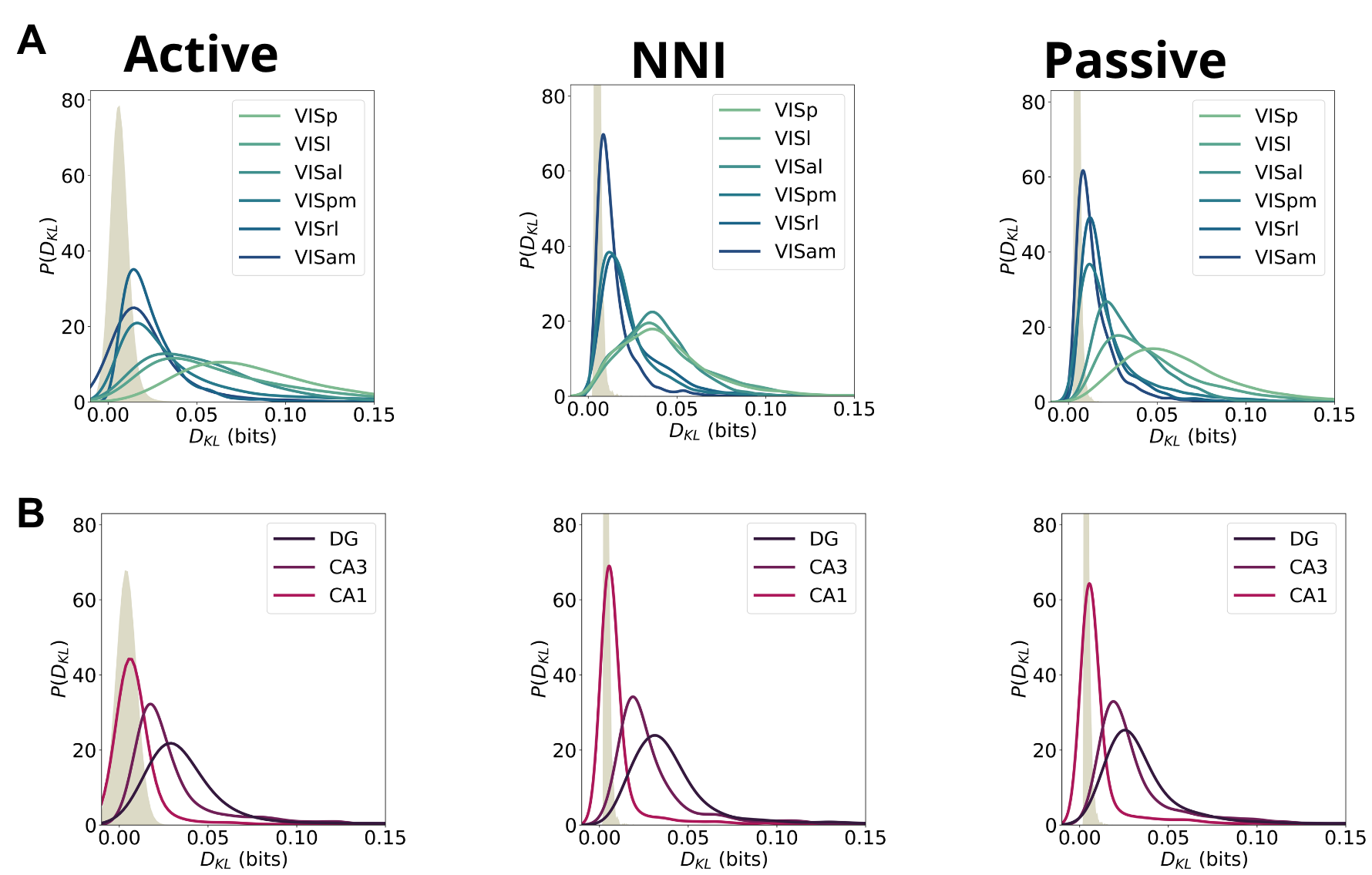}
		\caption{\textbf{Kullback-Lieber divergence between coarse-grained activity and Gaussian curve.} (A) and (B) display the distributions of the Kullback-Leibler divergence relative to a normal distribution ($D_{KL}$) in the visual cortex and hippocampus, respectively. Similar to the trends observed with $D_{JS}$, these values approach the trivial baseline as one moves deeper into the anatomical hierarchy.}
		\label{fig:kl_dists}
	\end{figure*}
	
	\begin{figure*}[ht!]
		\centering
		\includegraphics[width=\textwidth]{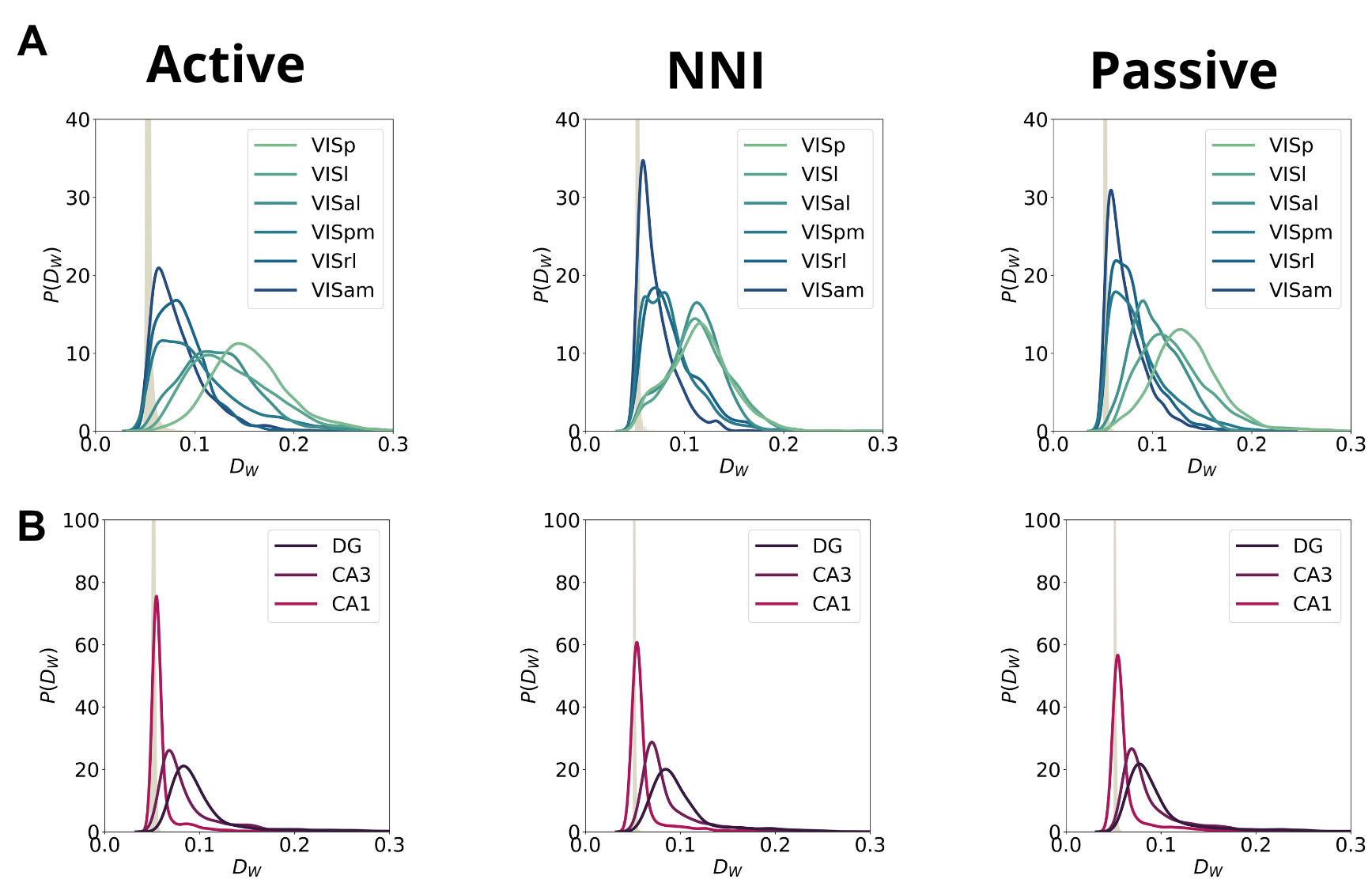}
		\caption{\textbf{Wasserstein distance between coarse-grained activity and a Gaussian baseline.} (A) and (B) display the distributions of the Wasserstein distance ($D_W$) in the visual cortex and hippocampus, respectively. Similar to the trends observed with $D_{JS}$, these values approach the trivial baseline as one moves deeper into the anatomical hierarchy. }
		\label{fig:wass_dists}
	\end{figure*}

	\begin{figure*}[ht!]
		\centering
		\includegraphics[width=\textwidth]{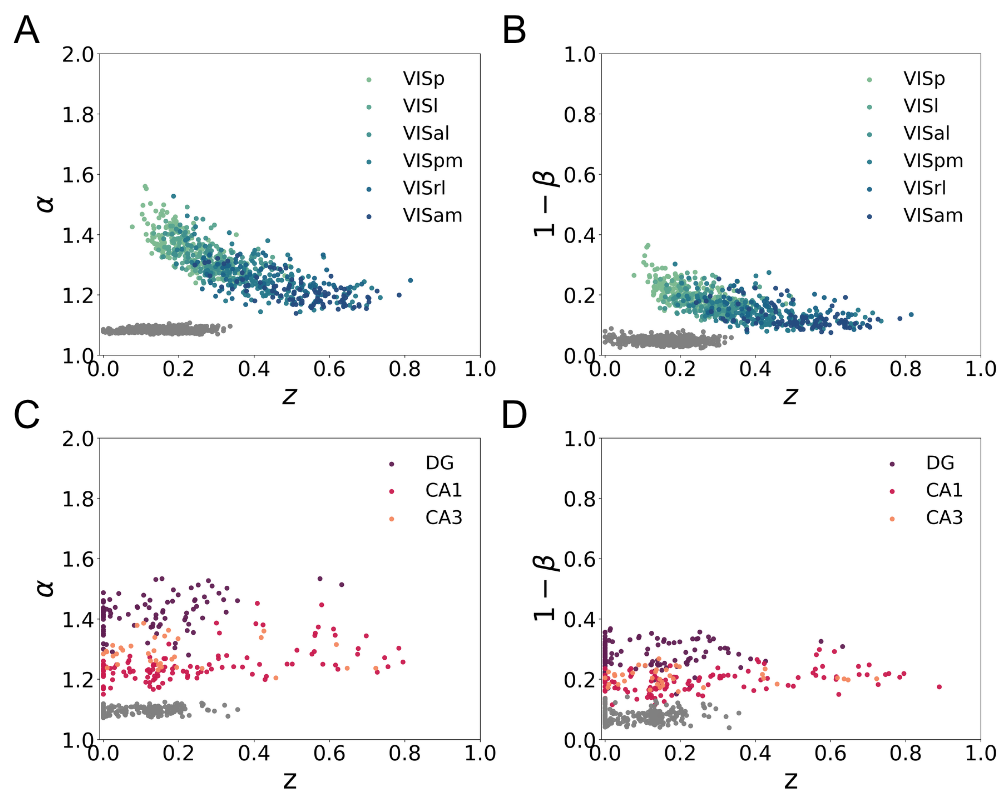}
		\caption{\textbf{Empirical relationship between static and dynamical scaling exponents.} Scatter plots displaying the relationship between the dynamical exponent $z$ (x-axis) and the static exponents $\alpha$ and $1-\beta$. (A) and (B) show the visual cortex, both static exponents exhibit a clear inverse coupling with $z$, reflecting the conflicting trends of static and dynamical criticality signatures along the anatomical hierarchy. (C, D) In the hippocampus, this coupling is flatter and less pronounced. Across all panels, gray markers indicate the exponents obtained from surrogate data, which collapse near the trivial fixed point values.}
		\label{fig:alpha_z}
	\end{figure*}
	
	\section*{Statistical Analysis}
	To compare the scaling exponents ($\alpha$, $\beta$, $z$) across regions within the visual cortex and hippocampus, we first performed a one-way ANOVA for each anatomical network (Table~I). We then conducted pairwise comparisons between regions within the same network. All pairwise $p$-values were adjusted using the Bonferroni correction to account for multiple comparisons. We report these adjusted $p$-values for all exponents across the experimental phases (Tables~II-IV), excluding cases where the initial ANOVA was $p \ge 0.05$.

	\begin{table}
		\centering
		
		\label{tab:anova_master}
		\begin{tabular}{|l|l|c|c|c|c|}
			\hline
			Experimental Phase &   Brain region & Variable &          p-value \\ \hline
			
			Active & Visual Cortex & $\alpha$ & $1.16 \times 10^{-38}$ \\ \hline
			Active & Visual Cortex &  $\beta$ & $1.12 \times 10^{-28}$ \\ \hline
			Active & Visual Cortex &      $z$ & $5.51 \times 10^{-41}$ \\ \hline
			Active &   Hippocampus & $\alpha$ & $4.18 \times 10^{-41}$ \\ \hline
			Active &   Hippocampus &  $\beta$ & $2.83 \times 10^{-21}$ \\ \hline
			Active &   Hippocampus &      $z$ &                  0.056 \\ \hline
			Passive & Visual Cortex & $\alpha$ & $5.63 \times 10^{-45}$ \\ \hline
			Passive & Visual Cortex &  $\beta$ & $3.99 \times 10^{-29}$ \\ \hline
			Passive & Visual Cortex &      $z$ & $1.72 \times 10^{-52}$ \\ \hline
			Passive &   Hippocampus & $\alpha$ & $3.97 \times 10^{-35}$ \\ \hline
			Passive &   Hippocampus &  $\beta$ & $1.02 \times 10^{-31}$ \\ \hline
			Passive &   Hippocampus &      $z$ &  $5.26 \times 10^{-5}$ \\ \hline
			NNI & Visual Cortex & $\alpha$ & $9.67 \times 10^{-39}$ \\ \hline
			NNI & Visual Cortex &  $\beta$ & $2.82 \times 10^{-29}$ \\ \hline
			NNI & Visual Cortex &      $z$ & $3.00 \times 10^{-50}$ \\ \hline
			NNI &   Hippocampus & $\alpha$ & $3.17 \times 10^{-35}$ \\ \hline
			NNI &   Hippocampus &  $\beta$ & $1.47 \times 10^{-26}$ \\ \hline
			NNI &   Hippocampus &      $z$ &                  0.007 \\ \hline
			
		\end{tabular}
		\caption{Global one-way ANOVA results for scaling exponents across anatomical networks.}
	\end{table}

	\begin{table}
		\centering
		
		\label{tab:stats_clean_naturalimages}
		\begin{tabular}{|l|c|c|c|}
			\hline
			Comparison &         $\alpha$ p-val &          $\beta$ p-val &              $z$ p-val \\ \hline
			VISp vs VISl &  $3.55 \times 10^{-4}$ &  $8.38 \times 10^{-6}$ &                  0.001 \\ \hline
			VISp vs VISal &  $1.63 \times 10^{-5}$ &  $1.55 \times 10^{-8}$ &  $2.35 \times 10^{-8}$ \\ \hline
			VISp vs VISpm & $5.62 \times 10^{-17}$ & $4.52 \times 10^{-15}$ & $6.12 \times 10^{-18}$ \\ \hline
			VISp vs VISrl & $1.02 \times 10^{-20}$ & $6.93 \times 10^{-16}$ & $3.73 \times 10^{-22}$ \\ \hline
			VISp vs VISam & $2.74 \times 10^{-25}$ & $1.41 \times 10^{-16}$ & $4.74 \times 10^{-23}$ \\ \hline
			VISl vs VISal &                   1.00 &                  0.314 &                  0.160 \\ \hline
			VISl vs VISpm &  $2.15 \times 10^{-7}$ &  $8.80 \times 10^{-5}$ &  $2.32 \times 10^{-9}$ \\ \hline
			VISl vs VISrl & $2.34 \times 10^{-12}$ &  $1.29 \times 10^{-7}$ & $4.78 \times 10^{-14}$ \\ \hline
			VISl vs VISam & $3.64 \times 10^{-16}$ &  $6.84 \times 10^{-8}$ & $1.04 \times 10^{-14}$ \\ \hline
			VISal vs VISpm &                  0.005 &                  0.460 &  $5.09 \times 10^{-4}$ \\ \hline
			VISal vs VISrl &  $1.99 \times 10^{-6}$ &                  0.008 &  $3.51 \times 10^{-8}$ \\ \hline
			VISal vs VISam &  $4.39 \times 10^{-9}$ &                  0.004 &  $1.53 \times 10^{-8}$ \\ \hline
			VISpm vs VISrl &                   1.00 &                   1.00 &                  0.027 \\ \hline
			VISpm vs VISam &                  0.054 &                   1.00 &                  0.012 \\ \hline
			VISrl vs VISam &                   1.00 &                   1.00 &                   1.00 \\ \hline
			DG vs CA3 & $1.62 \times 10^{-16}$ &  $7.27 \times 10^{-7}$ &                  0.616 \\ \hline
			DG vs CA1 & $1.44 \times 10^{-38}$ & $1.49 \times 10^{-19}$ &                  0.051 \\ \hline
			CA3 vs CA1 &                  0.004 &                  0.084 &                   1.00 \\ \hline
		\end{tabular}
		\caption{Bonferroni-adjusted p-values of pairwise comparison for the active experimental phase.}
	\end{table}
	
	\begin{table}
		\centering
		
		\label{tab:stats_clean_naturalreplay}
		\begin{tabular}{|l|c|c|c|}
			\hline
			Comparison &         $\alpha$ p-val &          $\beta$ p-val &              $z$ p-val \\ \hline
			VISp vs VISl &  $7.14 \times 10^{-4}$ &  $6.39 \times 10^{-6}$ & $7.98 \times 10^{-11}$ \\ \hline
			VISp vs VISal &  $7.70 \times 10^{-8}$ &  $4.94 \times 10^{-9}$ & $1.62 \times 10^{-13}$ \\ \hline
			VISp vs VISpm & $7.92 \times 10^{-23}$ & $4.08 \times 10^{-18}$ & $1.00 \times 10^{-27}$ \\ \hline
			VISp vs VISrl & $4.03 \times 10^{-15}$ &  $9.12 \times 10^{-9}$ & $3.32 \times 10^{-21}$ \\ \hline
			VISp vs VISam & $4.54 \times 10^{-27}$ & $2.10 \times 10^{-17}$ & $6.39 \times 10^{-33}$ \\ \hline
			VISl vs VISal &                  0.135 &                  0.357 &                  0.019 \\ \hline
			VISl vs VISpm & $8.12 \times 10^{-13}$ &  $7.48 \times 10^{-6}$ & $3.70 \times 10^{-15}$ \\ \hline
			VISl vs VISrl &  $5.56 \times 10^{-8}$ &                  0.118 & $2.58 \times 10^{-11}$ \\ \hline
			VISl vs VISam & $1.19 \times 10^{-18}$ &  $1.16 \times 10^{-7}$ & $9.18 \times 10^{-21}$ \\ \hline
			VISal vs VISpm &  $1.49 \times 10^{-5}$ &                  0.157 &  $1.12 \times 10^{-7}$ \\ \hline
			VISal vs VISrl &                  0.004 &                   1.00 &  $1.95 \times 10^{-5}$ \\ \hline
			VISal vs VISam & $3.78 \times 10^{-11}$ &                  0.003 & $1.55 \times 10^{-12}$ \\ \hline
			VISpm vs VISrl &                   1.00 &                   1.00 &                   1.00 \\ \hline
			VISpm vs VISam &                  0.106 &                   1.00 &                  0.023 \\ \hline
			VISrl vs VISam &                  0.004 &                  0.043 &                  0.115 \\ \hline
			DG vs CA3 & $1.24 \times 10^{-11}$ & $1.71 \times 10^{-10}$ &                  0.490 \\ \hline
			DG vs CA1 & $2.71 \times 10^{-32}$ & $7.35 \times 10^{-29}$ &  $2.34 \times 10^{-5}$ \\ \hline
			CA3 vs CA1 &  $1.46 \times 10^{-4}$ &                  0.012 &                  0.217 \\ \hline
		\end{tabular}
		\caption{Bonferroni-adjusted p-values of pairwise comparison for the passive experimental phase.}
	\end{table}
	
	\begin{table}
		\centering
		
		\label{tab:stats_clean_restexp}
		\begin{tabular}{|l|c|c|c|}
			\hline
			Comparison &         $\alpha$ p-val &          $\beta$ p-val &              $z$ p-val \\ \hline
			VISp vs VISl &                  0.009 &                  0.003 &                  0.064 \\ \hline
			VISp vs VISal &                   1.00 &  $3.12 \times 10^{-5}$ &  $6.86 \times 10^{-5}$ \\ \hline
			VISp vs VISpm & $1.40 \times 10^{-10}$ & $4.47 \times 10^{-19}$ & $1.38 \times 10^{-21}$ \\ \hline
			VISp vs VISrl &  $5.17 \times 10^{-7}$ &  $8.78 \times 10^{-5}$ & $1.66 \times 10^{-16}$ \\ \hline
			VISp vs VISam & $2.70 \times 10^{-23}$ & $1.19 \times 10^{-16}$ & $7.32 \times 10^{-35}$ \\ \hline
			VISl vs VISal &                  0.540 &                   1.00 &                  0.463 \\ \hline
			VISl vs VISpm & $2.05 \times 10^{-14}$ & $7.48 \times 10^{-10}$ & $1.06 \times 10^{-13}$ \\ \hline
			VISl vs VISrl & $7.72 \times 10^{-10}$ &                  0.495 & $1.38 \times 10^{-10}$ \\ \hline
			VISl vs VISam & $2.02 \times 10^{-24}$ & $2.21 \times 10^{-10}$ & $3.28 \times 10^{-26}$ \\ \hline
			VISal vs VISpm &  $1.12 \times 10^{-7}$ &  $1.20 \times 10^{-4}$ &  $3.37 \times 10^{-7}$ \\ \hline
			VISal vs VISrl &  $2.66 \times 10^{-5}$ &                   1.00 &  $1.22 \times 10^{-5}$ \\ \hline
			VISal vs VISam & $5.09 \times 10^{-17}$ &  $6.40 \times 10^{-6}$ & $1.41 \times 10^{-17}$ \\ \hline
			VISpm vs VISrl &                   1.00 &                  0.036 &                   1.00 \\ \hline
			VISpm vs VISam &  $3.65 \times 10^{-4}$ &                   1.00 &                  0.002 \\ \hline
			VISrl vs VISam &                  0.002 &                  0.005 &                  0.005 \\ \hline
			DG vs CA3 & $1.52 \times 10^{-12}$ &  $1.76 \times 10^{-9}$ &                  0.015 \\ \hline
			DG vs CA1 & $1.05 \times 10^{-32}$ & $2.45 \times 10^{-24}$ &                  0.012 \\ \hline
			CA3 vs CA1 &                  0.002 &                  0.158 &                   1.00 \\ \hline
		\end{tabular}
		\caption{Bonferroni-adjusted p-values of pairwise comparison for the NNI experimental phase.}
	\end{table}

\end{widetext}

\end{document}